\documentclass[preprint,sort&compress, 5p,times]{elsarticle}

\usepackage{lipsum}
\makeatletter
\def\ps@pprintTitle{%
 \let\@oddhead\@empty
 \let\@evenhead\@empty
 \def\@oddfoot{}%
 \let\@evenfoot\@oddfoot}
\makeatother

\usepackage{float,ulem}
\usepackage{graphicx,epsfig,epstopdf,amsmath,amssymb}
\usepackage{xcolor}
\usepackage[colorlinks=true,linkcolor=blue,citecolor=blue,urlcolor=
blue]{hyperref}

\newcommand{\be}{\begin{equation}}
\newcommand{\ee}{\end{equation}}
\newcommand{\ba}{\begin{eqnarray}}
\newcommand{\ea}{\end{eqnarray}}

\usepackage{caption}
\usepackage{subcaption}

\begin{document}
\title{An expedition to the islands of stability in  the first-order causal hydrodynamics}
%\title{Causality implies Lorentz invariance of stability in the first-order relativistic dissipative hydrodynamic theory}

\author{Rajesh Biswas}
\ead{rajeshbiswas@niser.ac.in}

\author{Sukanya Mitra}
\ead{sukanya.mitra@niser.ac.in}
%\affiliation{School of Physical Sciences, National Institute of Science Education and Research, An OCC of Homi Bhabha Nuclear Institute, Jatni-752050, India}

\author{Victor Roy}
\ead{victor@niser.ac.in}
\address{School of Physical Sciences, National Institute of Science Education and Research, An OCC of Homi Bhabha National Institute, Jatni-752050, India.}

\begin{abstract}
The recently proposed connection between the Lorentz 
invariance of stability and the speed of signal propagation has been tested for a first-order relativistic dissipative hydrodynamic theory. The fact that the stability situation in different reference frames 
agrees with each other only as long as the signal propagation respects causality, has been explicitly established for the theory, which is microscopically derived from the covariant kinetic equation in general 
hydrodynamic frames with arbitrary momentum-dependent interactions.
\end{abstract}
\maketitle

\section{Introduction}
Since the inception of the special theory of relativity~(STR), there has been a consensus that faster-than-light communication is not possible. To our best knowledge, it also has not been achieved in any experiment. However, several theories and phenomena related to superluminal communication have been proposed or studied, including tachyons \cite{Feinberg:1967}, neutrinos, quantum nonlocality, etc. The superluminal signal propagations are not permitted according to STR because, in a Lorentz-invariant theory, it could be used to transmit information into the past. This complicates causality leading to logical paradoxes of 
the ``kill your own grandfather" type. On the other hand, examples of matter of very high density allowing wave modes with superluminal group velocities have been known for a long time \cite{Bludman:1968}. 
The early formulations of the relativistic dissipative hydrodynamics (a.k.a the first-order hydrodynamic theories because the dissipative corrections to the energy-momentum tensor were constructed 
from the first-order gradient of fluid variables) by Eckart and Landau are also known to be inherently acausal due to the spacelike nature of the characteristics \cite{Hiscock:1985zz,Kostadt:2000ty}. 
%Since the inception of the special theory of relativity~(STR) there is a consensus that faster-than-light communication is not possible, and to date it has not been achieved in any experiment to our best knowledge. However, a number of theories and phenomena related to superluminal communication have been proposed or studied, including tachyons \cite{Feinberg:1967}, neutrinos, quantum nonlocality etc.  The superluminal signal propagations are not permitted according to STR because, in a Lorentz-invariant theory, it could be used to transmit information into the past. This complicates causality leading to logical paradoxes of the "kill your own grandfather" type. On the other hand, examples of matter of very high density allowing wave modes with superluminal group velocities are known for a long time Bludman, Ruderman. Particularly, the early formulations of the relativistic dissipative hydrodynamics theory (a.k.a the first-order hydrodynamic theories because the 
%dissipative corrections to the energy-momentum tensor of the fluid was constructed from the first-order gradient of fluid variables) by Eckart and Landau suffered from this type of pathology Ref[]. 
The acausality problem was later cured by Mueller-Israel-Stewarts~(MIS), where the entropy flux includes the higher order dissipative terms \cite{Israel:1979wp}.
At the same time, people also realized that the causality of relativistic hydrodynamics is closely associated with the stability of the fluid under small perturbations. Usually, any small perturbation
in fluid propagates with a characteristic speed of sound and transports energy from the source of the perturbation to distant points; the amplitude of the perturbation usually diminishes due to the dissipation of kinetic energy into heat which increases the total entropy, and the system reaches a new thermodynamic stable state. In some cases, the small perturbation may grow exponentially, eventually leading to instability in the fluid \cite{LLVol6}. 
Energy conservation, however, enforces that the instability cannot grow indefinitely unless the temperature of the system tends to infinity 
at the same time \cite{Hiscock:1988jd}. The situation for relativistic fluids became further complicated since it was found that a state of fluid can be stable to perturbations in one Lorentz frame and unstable in another, which clearly violates the principle of relativity.  
%In a recent work L. Gavssino et.al., provide a nice intuitive understanding of this puzzle.They show that two observers can disagree on whether a state is stable or unstable only if the perturbations can exit the light cone. They further showed that if a perturbation exits the forward light cone and its intensity changes with time due to dissipation, then there are instances where two observers disagree on the stability of the state. 
Hence one must make sure that any consistent relativistic hydrodynamic formulation must testify to the two benchmark criteria to be physically acceptable: (i) signal propagation must be subluminal to respect the causality, and (ii) the system must be stable in all Lorentz frames against any perturbations from the equilibrium state.The close correlation between them has been studied for MIS theory \cite{Hiscock:1983zz,Olson:1990rzl,Pu:2009fj}, where the theory is shown to be unstable because of superluminal modes. In the context of the recently proposed first-order stable-causal (BDNK) theory, 
\cite{Bemfica:2020zjp} gives a theorem that asserts that stability in the fluid's local rest frame implies stability in any Lorentz boosted frame, provided that the system is causal and strong hyperbolic.
Finally, in \cite{Gavassino:2021kjm}, the authors provide an intuitive understanding of this puzzle. They show that in presence of dissipation two observers can disagree on whether a state is stable or unstable iff the perturbations can exit the light cone. In conclusion, the stability of a state is a Lorentz-invariant property of dissipative theories iff the principle of causality holds.
%in \cite{Gavassino:2021kjm,Gavassino:2021owo} the correlation has been discussed in a more general set up where the first one shows that entropy maximization criteria of thermodynamic stability itself establishes causality, and the second one proves that stability is a Lorentz-invariant property of dissipative theories only if the perturbations remains inside the light cone. 
Motivated by these studies, we explicitly demonstrate here the said connection for a stable-causal first-order theory, derived from a first principle kinetic theory and valid in general hydrodynamic frames and with arbitrary momentum-dependent interactions.  
In \cite{Biswas:2022cla}, a first-order stable-causal relativistic theory (henceforth called BMR approach for brevity) has been derived, motivated from the works of \cite{BDNK} (BDNK theory) that presents a pathology
-free first-order relativistic theory via hydrodynamic field redefinition. The BMR approach showed that along with the general hydrodynamic frame choice the momentum dependence of microscopic interaction rate are also imperative for producing a causal and stable first-order relativistic theory. In the current work, we have checked that for BMR approach, all microscopic parameters considered to define the hydrodynamic frame and momentum dependence of the interaction rate, the linear stability analysis can differ from one Lorentz frame to another, if and only if the signal propagation exits the light cone. Stability is Lorentz invariant for all time-like perturbations; hence, testing it in one reference frame suffices for causal wave propagation. 
%This is a much-desired situation since stability in the fluid's local rest frame (which has a much-simplified dispersion relation polynomial) implies stability in any Lorentz-boosted frame, provided that the system is causal. 

Throughout the manuscript, we have used natural unit~($\hbar = c = k_{B} = 1 $) and flat space-time with mostly negative metric signature $g^{\mu\nu} = \text{diag}\left(1,-1,-1,-1\right)$.

\section{BMR theory}
In \cite{Biswas:2022cla}, a first-order, relativistic, stable and causal hydrodynamic theory has been derived in general frames from the Boltzmann transport equation, where the system interactions are introduced 
via the microscopic particle momenta captured through momentum-dependent relaxation time approximation (MDRTA) \cite{MDRTA}. The basic idea is to estimate the out-of-equilibrium one-particle distribution function 
$f(x,p)$ for general hydrodynamic frame and arbitrary momentum dependent interactions. $f(x,p)$ can be decomposed in two components : one is the homogeneous solution which sets the hydrodynamic frame choice and 
the second is the inhomogeneous solution that is controlled by the system interaction. To extract the homogeneous part of the solution, we use the matching conditions (constraints that set the thermodynamic fields  
to their equilibrium values even in the presence of dissipation). For estimating the inhomogeneous part, the relativistic transport equation, $p^{\mu}\partial_{\mu}f(x,p)=-\cal{L}[\phi]$, is solved where 
$\cal{L}[\phi]$ is the linearized collision operator with $\phi$ as the deviation in particle distribution due to dissipation. $p^{\mu}$ and $x$ denote the particle four-momenta and space-time variable respectively. 
We employ here the MDRTA method to explicitly linearize the collision term, with the relaxation time of single particle distribution function $\tau_R$ expressed as a power law of scaled particle energy 
($\tilde{E_p}=p^{\mu}u_{\mu}/T$) in comoving frame, $\tau_R=\tau_R^0(x)(\tilde{E_p})^{\Lambda}$. We propose an appropriate collision operator (that satisfies all the conservation properties, adjoint properties and 
summation invariant properties) for a semi-orthogonal, monic polynomial basis of the distribution function, instead of the Anderson-Witting type relaxation kernel, 
\begin{align}
{\cal{L}}_{\text{MDRTA}}[\phi]
&=\frac{\left(p\cdot u\right)}{\tau_R}f^{(0)}(1\pm f^{(0)})\bigg[\phi-
\tilde{p}_{\langle\nu\rangle}\frac{\langle\frac{\tilde{E}_p}{\tau_R}\phi \tilde{p}^{\langle\nu\rangle}\rangle}
{\frac{1}{3}\langle\frac{\tilde{E}_p}{\tau_R}\tilde{p}^{\langle\mu\rangle}\tilde{p}_{\langle\mu\rangle}\rangle}
\nonumber\\
&-\frac{\langle\frac{\tilde{E}_p}{\tau_R}\tilde{E}_p^2\rangle\langle\frac{\tilde{E}_p}{\tau_R}\phi\rangle-\langle\frac{\tilde{E}_p}{\tau_R}\tilde{E}_p\rangle\langle\frac{\tilde{E}_p}{\tau_R}\phi\tilde{E}_p\rangle}
{\langle\frac{\tilde{E}_p}{\tau_R}\rangle\langle\frac{\tilde{E}_p}{\tau_R}\tilde{E}_p^2\rangle-\langle\frac{\tilde{E}_p}{\tau_R}\tilde{E}_p\rangle^2}\nonumber\\
&-\tilde{E}_p\frac{\langle\frac{\tilde{E}_p}{\tau_R}\tilde{E}_p\rangle\langle\frac{\tilde{E}_p}{\tau_R}\phi\rangle-\langle\frac{\tilde{E}_p}{\tau_R}\rangle\langle\frac{\tilde{E}_p}{\tau_R}\phi\tilde{E}_p\rangle}
{\langle\frac{\tilde{E}_p}{\tau_R}\tilde{E}_p\rangle^2-\langle\frac{\tilde{E}_p}{\tau_R}\rangle\langle\frac{\tilde{E}_p}{\tau_R}\tilde{E}_p^2\rangle}\bigg]~.
\label{MDRTAcoll}
\end{align}
This collision operator preserves the conservation laws microscopically. By virtue of that, the right hand side of the transport equation singularly excludes the zero modes of the linearized collision operator.
Because of the fact, the left hand side of the transport equation is not necessarily needed to be orthogonal to zero modes and hence the covariant time derivatives appearing on the left hand side of the 
transport equation are not necessarily required to be exchanged by the spatial gradients. As a result, to solve the equation, we can safely apply a perturbative method keeping the time derivatives over the 
fundamental thermodynamic quantities alive (necessarily for a stable-causal theory) in the first-order thermodynamic field corrections as the following,
\begin{align}
\left[\begin{array}{l}
\delta n^{(1)} \\
\delta \epsilon^{(1)} \\
\delta P^{(1)}
\end{array}\right]&=\left[\begin{array}{l}
\nu_1 \\
\varepsilon_1 \\
\pi_1
\end{array}\right] \frac{D T}{T}+\left[\begin{array}{l}
\nu_2 \\
\varepsilon_2 \\
\pi_2
\end{array}\right] \left(\partial \cdot u \right)+\left[\begin{array}{l}
\nu_3 \\
\varepsilon_3 \\
\pi_3
\end{array}\right] D \tilde{\mu}~,
\label{sc}
\\
\left[\begin{array}{l}
 W^{(1)\mu} \\
V^{(1)\mu}
\end{array}\right]&=\left[\begin{array}{l}
\theta_{1} \\
\gamma_{1}
\end{array}\right] \left(\frac{\nabla^{\mu}T}{T} - Du^{\mu}\right)+\left[\begin{array}{l}
\theta_{3} \\
\gamma_{3}
\end{array}\right] \nabla^{\mu}\tilde{\mu}~.
\label{vec}
\end{align}

%%%%%%%%%%%%%%%%%%%
\iffalse
\begin{align}
&\delta n^{(1)}, \delta \epsilon^{(1)}, \delta P^{(1)}=\nu_{1},\varepsilon_{1},\pi_{1} \frac{DT}{T} + \nu_{2} ,\varepsilon_{2},\pi_{2}\left(\partial \cdot u\right)\nonumber\\
&~~~~~~~~~~~~~~~~~~~~~~+\nu_{3},\varepsilon_{3},\pi_{3} D\tilde{\mu}~,
\label{sc}\\
&W^{(1)\mu},V^{(1)\mu} = \theta_{1},\gamma_{1}\left[\frac{\nabla^{\mu}T}{T} - Du^{\mu}\right] +\theta_{3},\gamma_{3} \nabla^{\mu}\tilde{\mu}~.
\label{vec}
\end{align}
\fi
%%%%%%%%%%%%%%%%%%%%%%%%
Here, $\delta n, \delta\epsilon, \delta P, W^{\alpha}$ and $V^{\alpha}$ respectively denote dissipative corrections over particle number density, energy density, pressure, energy flow and particle current. The 
field corrections include first order derivatives over all the fundamental thermodynamic quantities such as temperature $T$, chemical potential $\mu$ and hydrodynamic four-velocity $u^{\mu}$. The field correction 
coefficients turn out to be elaborate functions of the frame indices $i,j,k$ and the parameter $\Lambda$ of MDRTA (for details see \cite{Biswas:2022cla}). These field corrections along with 
$\pi^{(1)\mu\nu}=2\eta\sigma^{\mu\nu}$ ($\eta$ is shear viscosity), constitute the first order out of equilibrium particle four-flow, $N^{\mu}=(n+\delta n^{(1)})u^{\mu}+V^{\mu(1)}$, and energy-momentum tensor,
$T^{\mu\nu}=\left(\epsilon+\delta\epsilon^{(1)}\right)u^{\mu}u^{\nu}-\left(P+\delta P^{(1)}\right)\Delta^{\mu\nu}+\left(W^{\mu(1)}u^{\nu}+W^{\nu}u^{\mu}\right)+\pi^{\mu\nu(1)}$, respectively.
Here, $n, \epsilon$ and $P$ indicate particle number density, energy density and pressure at their local equilibrium values. We observed that both the conventional frame choices (Landau ($i,j,k=1,2,1$)
and Eckart ($i,j,k=1,2,0$)) and the momentum independent $\tau_R$ ($\Lambda=0$) lead to an acausal and unstable theory.

\section{Stability and causality analysis with boosted background}
In \cite{Biswas:2022cla}, the stability and causality of BMR theory have been investigated (with an affirmation that the theory is indeed stable and causal with a critical dependence on the system interaction)
in the Lorentz rest frame of the fluid. But it has been observed a number of times \cite{Denicol:2008ha,Mitra:2021ubx} that such an analysis in local rest frame is often inadequate (even sometimes misleading) 
because with a boosted background the stability situation can alter drastically, even completely new modes can appear due to the degree of the dispersion polynomial changes.
We linearize the conservation equations for small perturbations of fluid variables around the hydrostatic equilibrium, $\psi(t,x)=\psi_0+\delta\psi(t,x)$,
with the fluctuations expressed in their plane wave solutions via a Fourier 
transformation $\delta\psi(t,x)\rightarrow e^{i(\omega t-kx)} \delta\psi(\omega,k)$, (subscript $0$ indicates global equilibrium).
The background fluid 
is now considered to be boosted along x-axis with a constant velocity $\textbf{v}$, $u^{\mu}_0=\gamma(1,\textbf{v},0,0)$ with $\gamma=1/\sqrt{1-\textbf{v}^2}$. The 
corresponding velocity fluctuation is 
$\delta u^{\mu}=(\gamma \textbf{v} \delta u^x,\gamma\delta u^{x},\delta u^y,\delta u^z)$ which again gives $u^{\mu}_0\delta u_{\mu}=0$ to maintain velocity normalization. The dispersion relation for linear
perturbations can be obtained in the boosted frame by giving the transformations, $\omega \rightarrow \gamma(\omega - k \textbf{v})$ and $k^2\rightarrow \gamma^2(\omega-k\textbf{v})^2-\omega^2+k^2$ to the local 
rest frame. The resulting dispersion relation polynomial turns out to be too mathematically cumbersome (as expected) and it is only possible to extract the modes in limiting situations. The transverse or shear channel 
at low wave number limit $(k\rightarrow 0)$ gives only one non-hydrodynamic (gapped) frequency mode,
\be
\omega^{\perp}=i\frac{(\epsilon_0+P_0)}{\gamma(\theta-\eta \textbf{v}^2)}+{\cal{O}}(k)~,
\ee
where we define $\theta=-\theta_1$. The other one is a hydrodynamic (gapless) mode that vanishes at spatially homogeneous limit ($k=0$). Hence, from the criteria that the imaginary part of frequency
must be positive definite to give rise to exponentially decaying perturbations, the stability condition for the shear channel becomes $\theta>\eta \textbf{v}^2$. Here, we recall the asymptotic causality condition
for the shear channel in Lorentz rest frame (LRF) was $\theta>\eta$ \cite{Biswas:2022cla} which readily reproduces the stability condition for all $0<\textbf{v}<1$. With boosted background, the group velocity of the 
propagating shear mode becomes $v_g^{\perp}=\lim_{k\rightarrow\infty}\bigg\vert\frac{\partial \textrm{~Re}({\omega^{\perp}})}{\partial k}\bigg\vert=(\textbf{v}+\sqrt{\eta/\theta})/(1+\textbf{v}\sqrt{\eta/\theta})$, 
which is subluminal as long as the asymptotic causality condition in LRF is satisfied. 
%This is any way straightforward from the principle of relativity that if two events are timelike, they remain so for any  reference frame and consequently the associated signal will be subluminal in any Lorentz frame.

\begin{figure}
\includegraphics[width=0.45\textwidth]{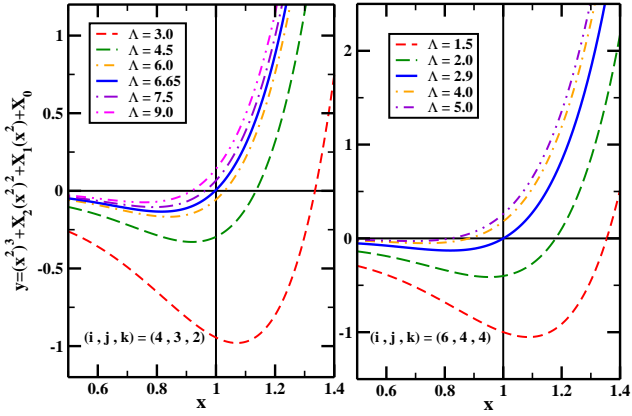} 
\caption{Estimating the value of $v_g^{\parallel}$ at LRF with different $\Lambda$ and in different frame situations. $\textbf{x}=1$ corresponds to the speed of light in vacuum.}
\label{vg}
\end{figure}

In longitudinal or sound channel the situation becomes quite involved since the number of modes significantly increases. At the asymptotic limit $k\rightarrow\infty$, the dispersion relation becomes,
\begin{align}
&H_6 (\omega^{\parallel})^6+H_5 (\omega^{\parallel})^5 k+H_4 (\omega^{\parallel})^4k^2+H_3 (\omega^{\parallel})^3 k^3\nonumber\\
&+H_2 (\omega^{\parallel})^2 k^4+H_1 (\omega^{\parallel})k^5+H_0k^6=0~.
\end{align}
The coefficients $H_{0,\cdots, 6}$ are elaborate functions of frame indices $i,j,k$, the MDRTA parameter $\Lambda$ and the boost velocity $\textbf{v}$. We will give here the explicit expressions of only those 
coefficients which are exclusively required for establishing causality and stability. In the limit of large $k$, an expansion of the form, $\omega^{\parallel}=v_g^{\parallel}k+\sum_{n=0}^{\infty}c_nk^{-n}$ 
can be used as a solution~ \cite{Brito:2020nou}. The values of the asymptotic group velocity $v_g^{\parallel}$ are then given by the sixth order polynomial,
$H_6(v_g^{\parallel})^6+H_5(v_g^{\parallel})^5+H_4(v^{\parallel}_g)^4+H_3(v^{\parallel}_g)^3+H_2(v_g^{\parallel})^2+H_1v_g^{\parallel}+H_0=0.$
Here, for simplicity, in Fig.(\ref{vg}), we extract the value of $v_g^{\parallel}$ in LRF for two different general frames, ($i,j,k$)=($4,3,2$) and ($i,j,k$)=($6,4,4$), for which the sixth order
polynomial reduces to a cubic equation, $(\textbf{x}^2)^3+X_2(\textbf{x}^2)^2+X_1(\textbf{x}^2)+X_0=0$ with $X_2=-A_4^2/A_6, X_1=A_2^4/A_6, X_0=-A_0^6/A_6$, whose roots will give the values of 
$v_g^{\parallel}$. Now, if the discriminant of this equation, $\Delta(=18X_2X_1X-0-4X_2^3X_0+X_2^2 X_1^2-4X_1^3-27X_0^2)>0$, we will have three distinct real roots, while for $\Delta<0$, we have only one real root 
and two complex conjugate roots. For small values of $\Lambda$, ($\Lambda<1.2$ for $(i,j,k)=(4,3,2)$ and $\Lambda<0.6$ for $(i,j,k)=(6,4,4)$), we get three real roots of $\textbf{x}^2$, among which two are always 
negative and one is always greater than one. So, that range fully excludes any subluminal root for $v_g^{\parallel}$. For the range of $\Lambda$ shown in Fig.(\ref{vg}), we have only one real, positive root for both the cases. We observe that increasing $\Lambda $ 
corresponds to smaller
$v_g^{\parallel}$ that eventually becomes subluminal. It can be seen, that the critical value of $\Lambda(=\Lambda_c)$ for reaching the asymptotic causality condition is, $\Lambda_c=6.65$ for frame $(i,j,k)=(4,3,2)$ and $\Lambda_c=2.9$ for frame $(i,j,k)=(6,4,4)$.
Hence, the theory is causal for $\Lambda > \Lambda_c$. So we conclude, that more general the frame situation becomes (farther away from Landau or Eckart frame where we can not 
find any subluminal $v_g$ for any $\Lambda$ value), the system reaches causality with smaller values of $\Lambda$. But that never reaches $\Lambda=0$ (momentum independent RTA) whatever large frame indices are.
It can be shown, that for the causal parameter sets shown in Fig.(\ref{vg}), the theory is asymptotically causal for any boosted frame as well, i.e, the original sixth order polynomial always has a subluminal root,
as suggested by the principle of relativity.

\begin{figure}
\includegraphics[width=0.45\textwidth]{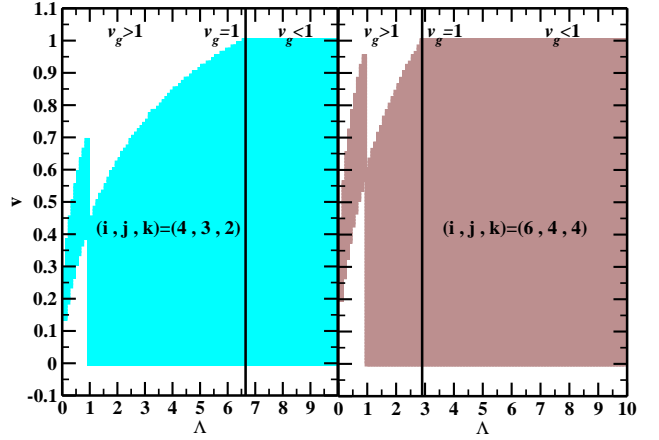}
\caption{Island of stability for different boost velocities and different $\Lambda$ values.}
\label{figstab}
\end{figure}
Next we turn to the stability analysis of the longitudinal modes. The dispersion relation at $k\rightarrow 0$ limit turns out to be,
\begin{align}
 &G_6 (i\omega ^{\parallel})^6 +G_5 (i\omega^{\parallel})^5 +G_4 (i\omega^{\parallel})^4 +G_3(i\omega^{\parallel})^ 3 +\nonumber\\
 &G_2 (i\omega^{\parallel})^2 (ik) +G_1 (i\omega^{\parallel}) k^2 +G_0 ik^3 = 0 ~.
 \label{Eqstab1}
\end{align}
Eq.\eqref{Eqstab1} gives rise to three hydrodynamic and three nonhydrodynamic modes. The hydro modes are given by, 
$\omega^{\parallel}=(a_{1,2,3})k$, where the group velocities can be estimated from the equation, $G_3 a^3+G_2 a^2-G_1 a-G_0=0$. The non-hydro modes (growing ones give rise to instability) are given by the following
polynomial equation, 
\be
G_6 (\omega^{\parallel})^3-i G_5 (\omega^{\parallel})^2 - G_4 \omega^{\parallel}+iG_3=0~.
\label{R-H}
\ee
Using Routh-Hurwitz stability criteria, we find the following conditions for stability of the non-hydro modes,
\be
G_6,G_5,G_3>0~,G_{B2}=(G_4 G_5-G_3 G_6)/G_5>0~.
\label{stabcond}
\ee
The coefficients are given by, 
$G_6=\gamma^6(A_6-A_4^2\textbf{v}^2+A_2^4\textbf{v}^4-A^6_0\textbf{v}^6),~ G_5=\gamma^5(A_5-A_3^2\textbf{v}^2+A_1^4\textbf{v}^4),~ G_4=\gamma^4(A_4^0-A_2^2\textbf{v}^2+A_0^4\textbf{v}^4)$ and 
$G_3=\gamma^3(A_3^0-A_1^2\textbf{v}^2)$. The explicit mathematical expressions of the $A$ functions in terms of ($i,j,k$) and $\Lambda$, through which the frame situation and the momentum 
dependence of the interaction cross section duly control the stability and causality of the theory, are given in the Appendix.

Among the coefficients of \eqref{stabcond}, $G_3=\gamma^3n_0(\epsilon_0+P_0)(1-\textbf{v}^2c_s^2)$ is always positive where $c_s$ is the velocity of sound. In the parameter space spanned by
($i,j,k$), $\Lambda$ and $\textbf{v}$, the positivity of $G_6, G_5$ and $G_{B2}$ ensures the stability of the system. In Fig.(\ref{figstab}), system stability has been displayed for two frame conditions,
$(i,j,k)=(4,3,2)$ and $(i,j,k)=(6,4,4)$, as a function of two independent parameters $\Lambda$ and $\textbf{v}$. Here, the shaded areas indicate that the system stability holds for those parameters
sets. Here, we notice something very interesting. The $\Lambda$ values for each frame, beyond which the asymptotic causality condition holds, $v_g<1$, ($\Lambda_c=6.65$ for $(i,j,k)=(4,3,2)$ and 
$\Lambda_c=2.9$ for $(i,j,k)=(6,4,4)$, see Fig.(\ref{vg})), the system is also stable irrespective of the values of boost velocity as long as $0<\textbf{v}<1$. This is the main result of this work.    
In the region where $v_g>1$, we observe that there is no $\Lambda$ value for which the system is either stable or unstable for all values of $\textbf{v}$, whereas beyond the $v_g=1$ line, the stability
situation agrees in all reference frames.

We now analyse the stability situation at large wave-number limit.
With a general boost velocity $\textbf{v}$, at large $k$, the shear channel dispersion relation gives the following modes,
\begin{equation}
\omega^{\perp}_{1,2}= k \frac{\bigg(\textbf{v}\pm \sqrt{\frac{\eta}{\theta}}\bigg)}{\bigg(1\pm \textbf{v}\sqrt{\frac{\eta}{\theta}}\bigg)} 
+ i \frac{(\epsilon_0+P_0)\bigg[1\pm \sqrt{\frac{\eta \textbf{v}^2}{\theta}}\bigg]}{2\gamma (\theta - \eta \textbf{v}^2)} + {\cal{O}}(\frac{1}{k})~.
\end{equation}
Clearly, with $0<\textbf{v}<1$ and asymptotic causality condition $\theta>\eta$, the imaginary part is always positive which ensures stability of the shear modes.
The situation at sound channel becomes somewhat non-trivial. With general $k$ and $\textbf{v}$, it is a sixth-order polynomial of $\omega$, with each coefficient itself as a polynomial over $k$, where 
the coefficient of each power of $k$ 
is again a polynomial of $\textbf{v}$. This is only possible to solve at limiting situations. Hence, we present here the solution upto the order of $k$, i.e, $\omega^{\parallel}=a_0+a_1 k$. $a_0$ is purely imaginary
with three roots whose positivity has been checked by Routh-Hurwitz stability criteria in Eq.\eqref{R-H}. We have checked that $a_1$ is a real quantity which does not contribute to the stability of the theory.
In view of the non-trivial structure of the sound channel polynomial, we believe that this result might suffice for a long-wavelength effective theory such as relativistic hydrodynamics.
It is in principle important to analyse also the larger $k$ limit, but in practice it is beyond the possibilities of the current work.

So, from this analysis we can safely conclude here that stability is a Lorentz invariant property if and only if the signal propagation within the medium
is causal. Our results corroborates with the finding of \cite{Gavassino:2021kjm}, where the authors addressed the physical 
origin of the relation between ``causality" and ``instability" using space-like events connected via perturbation mediating through a dissipative medium.
%In \cite{Gavassino:2021kjm}, a physical picture of this phenomena has been addressed recently. If the signal travels between two events that are spacelike separated (outside thelightcone), then by relativity of simultaneity we can always find some reference frame where the order of the events can be flipped, i,e, effect precedes the cause. In such a situation, a decaying perturbation in the dissipative medium in between the events, may appear spontaneously generated and growing without any external influence. This situation clearly implies stability within the system and prevents an acausal dissipative systems to be covariantly stable. 
Fig.(\ref{figstab}) gives a very clear pictorial representation of similar argument for the BMR theory, where we have established that for a first order relativistic dissipative system in general frame and with momentum dependent interaction cross section, the principle of causality is requisite for the stability
to be a Lorentz invariant property.

\section{Conclusion}
%The counterintuitive observation, that a relativistic fluid can be stable under perturbation in one inertial reference frame but not in others, needs to be resolved for the sake of a valid physical picture itself. 
The idea is that, causality violation can chronologically reorder a perturbation in different reference frames by the relativity of simultaneity so that in a dissipative medium, two observers can disagree on whether the perturbation is growing or decaying. This gives rise to instability in the medium, so causality is the key signature in this paradox. This phenomenon has been explicitly demonstrated in this work,
where all the stability controlling factors, such as frame indices and interaction parameters, give rise to a stable system at all inertial frames iff they respect the asymptotic causality for the system. This undeniably reduces the task of checking stability in all the reference frames, and the local rest frame suffices for the analysis. This is a much desirable situation since, with a boosted background, the dispersion relation becomes way more complicated than in the local rest frame. Most importantly, this work singularly establishes the physical argument
and the associated theorems for a coarse-grained system with momentum-dependent microscopic interactions from the first principle calculation.
The analysis of causality in this work is based on the asymptotic causality condition, which states that the asymptotic group velocity
$v_g=\lim _{k \rightarrow+\infty} \frac{d \operatorname{Re} \omega(k)}{d k}$ must be subluminal (i.e. $v_g \leq 1$). However, there are examples of acausal theories with subluminal group velocity, 
such as the diffusion equation: $\partial_t T=\partial_x^2 T$, with dispersion relation $\omega(k)=i k^2$ and $v_g=0 \leq 1$. And yet, we know that the 
diffusion equation is acausal. We note that the ``asymptotic causality condition'' is a necessary condition for causality, but not sufficient \cite{Krotscheck:1978}. A rigorous study of causality requires the study 
of characteristics and ensuring the strong hyperbolicity condition, which is beyond the scope of the present work.

\section{Appendix}
The $A$ functions mentioned in the main text are given by :
\begin{align}
&A_6=F_{13}(F_3F_6-F_1F_8)~,\\
&A_5=F_1F_8-F_3F_6-(\epsilon_0+P_0)F_1F_{13}-n_0F_8F_{13}~,\\
&A_4^0=(\epsilon_0+P_0)F_1+n_0F_8-n_0(\epsilon_0+P_0)F_{13}~,\\
&A_4^2=F_{13}(F_4F_6-F_2F_8)+F_5(F_8F_{11}-F_6F_{12})\nonumber\\
&~~~~-F_1(F_9F_{13}-F_8F_{14}-F_{10}F_{12})\nonumber\\
&~~~~-F_3(F_{10}F_{11}+F_6F_{14}-F_7F_{13})\\
&A_3^0=n_0(\epsilon_0+P_0)~,\\
&A_3^2=n_0(F_8F_{14}+F_{10}F_{12}-F_{9}F_{13}+F_{8}F_{11}-F_{6}F_{12})\nonumber\\
&+(\epsilon_0+P_0)(F_1F_{14}+F_{1}F_{12}-F_{2}F_{13}-F_{3}F_{11}+F_{5}F_{11})\nonumber\\
&+F_1F_9+F_2F_8-F_3F_7-F_4F_6\nonumber\\
&+b(F_1F_{10}-F_5F_6)+a(F_5F_8-F_3F_{10})~,\\
&A_2^2=(\epsilon_0+P_0)(F_1b+F_2-aF_3+aF_5)\nonumber\\
&~~~~+n_0(F_9+bF_{10}-bF_6+aF_8)\nonumber\\
&~~~~+n_0(\epsilon_0+P_0)(F_{11}+F_{12}+F_{14})~,\\
&A_2^4=F_2(F_8F_{14}+F_{10}F_{12}-F_9F_{13})\nonumber\\
&~~~~+F_4(F_7F_{13}-F_{10}F_{11}-F_6F_{14})\nonumber\\
&~~~~+F_5(F_9F_{11}-F_7F_{12})+F_{14}(F_1F_9-F_3F_7)~,\\
&A_1^2=n_0(\epsilon_0+P_0)(a+b)~,\\
&A_1^4=(F_2F_9-F_4F_7)+n_0(F_9F_{14}+F_9F_{11}-F_7F_{12})\nonumber\\
&~~~~+(\epsilon_0+p_0)(F_2F_{12}+F_2F_{14}-F_4F_{11})\nonumber\\
&~~~~+a(F_5F_9-F_4F_{10})+b(F_2F_{10}-F_5F_7)~,\\
&A_0^4=(\epsilon_0+P_0)(bF_2-aF_4)+n_0(aF_9-bF_7)~,\\
&A_0^6=F_{14}(F_2F_9-F_4F_7)~.
\end{align}
These $A$'s are the functions of the first order field correction coefficients (given in Eq.(2) and (3) of the main text) as the following,
\begin{align}
&F_1=\nu_1 f+\nu_3 c~,~~~~~F_2=\gamma_1 f+\gamma_3 c~,\\
&F_3=\nu_1 g+\nu_3 d~,~~~~~F_4=\gamma_1 g+\gamma_3 d~,\\
&F_5=\nu_2-\gamma_1~,~~~~~~~~F_6=\varepsilon_1 f+\varepsilon_3 c~,\\
&F_7=\theta_1 f+\theta_3 c~,~~~~~F_8=\varepsilon_1 g+\varepsilon_3 d~,\\
&F_9=\theta_1 g+\theta_3 d~,~~~~~F_{10}=\varepsilon_2-\theta_1~,\\
&F_{11}=[f(\pi_1-\theta_1)+c(\pi_3-\theta_3)]/(\epsilon_0+P_0)~,\\
&F_{12}=[g(\pi_1-\theta_1)+d(\pi_3-\theta_3)]/(\epsilon_0+P_0)~,\\
&F_{13}=\theta_1/(\epsilon_0+P_0)~,\\
&F_{14}=(4\eta/3)/(\epsilon_0+P_0)-\pi_2/(\epsilon_0+P_0)~.
\end{align}
Here, we have used, $a=\frac{n_0}{(\epsilon_0+P_0)}(\frac{\partial P}{\partial n})_{\epsilon}~,~b=(\frac{\partial P}{\partial\epsilon})_{n}~,~ 
c=J_0 I_3/(I_2^2 - I_1 I_3)~,~d=-J_1 I_2/(I_2^2 - I_1 I_3)~,~f=-J_0 I_2/(I_2^2 - I_1 I_3)$ and $g=J_1 I_1/(I_2^2 - I_1 I_3)$. We readily identify that, $c_s^2=a+b$. The moment integrals used here 
are defined as,
$I_{n}=\int dF_p \tilde{E}_{p}^n,~\Delta^{\mu\nu}J_n=\int dF_p \tilde{p}^{\langle\mu\rangle}\tilde{p}^{\langle\nu\rangle}\tilde{E}_{p}^n$, and
$\Delta^{\alpha\beta\mu\nu}K_n=\int dF_p\tilde{p}^{\langle\mu}\tilde{p}^{\nu\rangle}\tilde{p}^{\langle\alpha}\tilde{p}^{\beta\rangle}\tilde{E}_{p}^n$.
$z=m/T$ is the scaled rest mass of the particles.

Next, we give the explicit expressions of the first order field correction coefficients (scaled by $\tau_R^0$) in terms of the frame indices ($i,j,k$) and MDRTA parameter $\Lambda$, using the notation 
$\mathcal{D}_{i,j}^{m,n}=I_{i+m}I_{j+n}-I_{i+n}I_{j+m}$ :
%\begin{widetext}
\begin{align}
&\nu_1=-TI_{\Lambda+2}+\frac{\partial n}{\partial\tilde{\mu}}\frac{{\cal{D}}_{i,j}^{\Lambda+1,1}}{{\cal{D}}_{i,j}^{0,1}}+T\frac{\partial n}{\partial T}\frac{{\cal{D}}_{i,j}^{\Lambda+1,0}}{{\cal{D}}_{i,j}^{1,0}}~,\\
&\nu_2=-T\left[\frac{1}{3}I_{\Lambda+2}-\frac{z^2}{3}I_{\Lambda}\right]
+\frac{\partial n}{\partial\tilde{\mu}}\left[\frac{1}{3}\frac{{\cal{D}}_{i,j}^{\Lambda+1,1}}{{\cal{D}}_{i,j}^{0,1}}-\frac{z^2}{3}\frac{{\cal{D}}_{i,j}^{\Lambda-1,1}}{{\cal{D}}_{i,j}^{0,1}}\right] \nonumber \\
&~~~~+T\frac{\partial n}{\partial T}\left[\frac{1}{3}\frac{{\cal{D}}_{i,j}^{\Lambda+1,0}}{{\cal{D}}_{i,j}^{1,0}}-\frac{z^2}{3}\frac{{\cal{D}}_{i,j}^{\Lambda-1,0}}{{\cal{D}}_{i,j}^{1,0}}\right]~,\\
&\nu_3=-TI_{\Lambda+1}+\frac{\partial n}{\partial\tilde{\mu}}\frac{{\cal{D}}_{i,j}^{\Lambda,1}}{{\cal{D}}_{i,j}^{0,1}}+T\frac{\partial n}{\partial T}\frac{{\cal{D}}_{i,j}^{\Lambda,0}}{{\cal{D}}_{i,j}^{1,0}}~,\\
&\varepsilon_1=-T^2I_{\Lambda+3}+\frac{\partial\epsilon}{\partial\tilde{\mu}}\frac{{\cal{D}}_{i,j}^{\Lambda+1,1}}{{\cal{D}}_{i,j}^{0,1}}
+T\frac{\partial\epsilon}{\partial T}\frac{{\cal{D}}_{i,j}^{\Lambda+1,0}}{{\cal{D}}_{i,j}^{1,0}}~,\\
\nonumber
&\varepsilon_2=-T^2\left[\frac{1}{3}I_{\Lambda+3}-\frac{z^2}{3}I_{\Lambda+1}\right]
+\frac{\partial \epsilon}{\partial\tilde{\mu}}\left[\frac{1}{3}\frac{{\cal{D}}_{i,j}^{\Lambda+1,1}}{{\cal{D}}_{i,j}^{0,1}}-\frac{z^2}{3}\frac{{\cal{D}}_{i,j}^{\Lambda-1,1}}{{\cal{D}}_{i,j}^{0,1}}\right] \\
&~~~~+T\frac{\partial \epsilon}{\partial T}\left[\frac{1}{3}\frac{{\cal{D}}_{i,j}^{\Lambda+1,0}}{{\cal{D}}_{i,j}^{1,0}}-\frac{z^2}{3}\frac{{\cal{D}}_{i,j}^{\Lambda-1,0}}{{\cal{D}}_{i,j}^{1,0}}\right]~,\\
&\varepsilon_3=-T^2I_{\Lambda+2}+\frac{\partial\epsilon}{\partial\tilde{\mu}}\frac{{\cal{D}}_{i,j}^{\Lambda,1}}{{\cal{D}}_{i,j}^{0,1}}+
T\frac{\partial\epsilon}{\partial T}\frac{{\cal{D}}_{i,j}^{\Lambda,0}}{{\cal{D}}_{i,j}^{1,0}}~,\\
&\pi_1=-T^2\left[\frac{1}{3}I_{\Lambda+3}-\frac{z^2}{3}I_{\Lambda+1}\right]+\frac{\partial P}{\partial\tilde{\mu}}\frac{{\cal{D}}_{i,j}^{\Lambda+1,1}}{{\cal{D}}_{i,j}^{0,1}}
+T\frac{\partial P}{\partial T}\frac{{\cal{D}}_{i,j}^{\Lambda+1,0}}{{\cal{D}}_{i,j}^{1,0}}~,\\
&\pi_2=-T^2\left[\frac{1}{9}I_{\Lambda+3}-\frac{2}{9}z^2I_{\Lambda+1}+\frac{1}{9}z^4I_{\Lambda-1}\right] \nonumber \\ 
&~~~~+\frac{\partial P}{\partial\tilde{\mu}}\left[\frac{1}{3}\frac{{\cal{D}}_{i,j}^{\Lambda+1,1}}{{\cal{D}}_{i,j}^{0,1}}-\frac{z^2}{3}\frac{{\cal{D}}_{i,j}^{\Lambda-1,1}}{{\cal{D}}_{i,j}^{0,1}}\right] \nonumber \\
&~~~~+T\frac{\partial P}{\partial T}\left[\frac{1}{3}\frac{{\cal{D}}_{i,j}^{\Lambda+1,0}}{{\cal{D}}_{i,j}^{1,0}}-\frac{z^2}{3}\frac{{\cal{D}}_{i,j}^{\Lambda-1,0}}{{\cal{D}}_{i,j}^{1,0}}\right]~,\\
&\pi_3=-T^2\left[\frac{1}{3}I_{\Lambda+2}-\frac{z^2}{3}I_{\Lambda}\right]
+\frac{\partial P}{\partial\tilde{\mu}}\frac{{\cal{D}}_{i,j}^{\Lambda,1}}{{\cal{D}}_{i,j}^{0,1}}
+T\frac{\partial P}{\partial T}\frac{{\cal{D}}_{i,j}^{\Lambda,0}}{{\cal{D}}_{i,j}^{1,0}}~,\\
&\theta_1=- T^2 \left[ J_{\Lambda+1}+\frac{\epsilon_0+P_0}{T^2}\frac{J_{k+\Lambda}}{J_k}\right]~,\\
&\theta_3=- T^2 \left[ J_{\Lambda}+\frac{\epsilon_0+P_0}{T^2}\frac{J_{k+\Lambda-1}}{J_k}\right]~,\\
&\gamma_1=- T \left[ J_{\Lambda}+\frac{n_0}{T}\frac{J_{k+\Lambda}}{J_k}\right]~,\\
&\gamma_3=- T \left[ J_{\Lambda-1}+\frac{n_0}{T}\frac{J_{k+\Lambda-1}}{J_k}\right]~,\\
&\eta=\frac{1}{2}T^2 K_{\Lambda-1}~.
\end{align}
%\end{widetext}
%\vspace{15mm}

These $14$ first order transport coefficients, through their dependence on frame parameters and interaction parameter, critically control the stability and causality analysis presented in the 
current work. For the analysis we have used $T=300$ MeV and $m=300$ MeV.

\textit{Acknowledgements.}$\textendash$ 
R.B. acknowledges the financial support from SPS, NISER planned project RIN4001. S.M and V.R. acknowledges the financial support from  the Department of Atomic Energy,India.

\end{document}